\documentclass[10pt,letterpaper]{article}

\usepackage{cogsci}
\usepackage{graphicx}
\usepackage{booktabs}
\usepackage{hyperref}

\cogscifinalcopy 

\usepackage[
  style=apa,
  natbib=true,
  annotation=false,
]{biblatex}
\usepackage{float}

\title{Do Melody and Rhythm Coevolve?}

\author[1]{\mbox{Harin Lee}}
\author[2]{\mbox{Rainer Polak}}
\author[3]{\mbox{Manuel Anglada-Tort}}
\author[4]{\mbox{Marc Schönwiesner}}
\author[5]{\mbox{Minsu Park}}
\author[6]{\mbox{Nori Jacoby}}
\affil[1]{University of Cambridge, UK}
\affil[2]{RITMO Centre for Interdisciplinary Studies in Rhythm, Time and Motion, University of Oslo, Norway}
\affil[3]{Department of Psychology, Goldsmiths College, University of London, UK}
\affil[4]{Department of Life Sciences, Leipzig University, Germany}
\affil[5]{Division of Social Science, New York University Abu Dhabi, UAE}
\affil[6]{Department of Psychology, Cornell University, USA}

\begin{document}

\maketitle

\begin{abstract}
Music comprises two core structural components, melody and rhythm, that vary widely across cultures. Whether these components coevolve in a coupled way or follow independent trajectories remains unclear. We introduce a novel computational pipeline to extract vocal melodic pitch-interval and percussive inter-onset timing distributions from 27,628 popular songs across 59 countries, enabling large-scale cross-cultural comparison that bypasses traditional music annotations. Musical similarities between countries aligned with geographic and linguistic relationships, validating our approach. Substantial variation emerged in both melodic and rhythmic structures across countries, yet the diversity of the two components was not significantly correlated, challenging assumptions of coupled evolution. Only rhythmic diversity was significantly associated with ethnic and linguistic heterogeneity, while melodic diversity showed no such association. These findings suggest that melody and rhythm constitute partially independent systems shaped by distinct cultural and evolutionary pressures, rather than components of a single monolithic musical style.


\end{abstract}

\section{Introduction}\label{sec:introduction}
Songs are present in all human cultures and serve important social and individual functions~\citep{mehr_universality_2019, savage_statistical_2015}. Across cultures, songs consistently comprise two core structural components: melody, which organizes pitch, and rhythm, which structures temporal durations~\citep{mehr_core_2025}. Both components show cross-cultural regularities consistent with shared cognitive constraints, yet they also vary widely both within and between societies, reflecting culturally learned traditions that change over time~\citep{jacoby_commonality_2024, ozaki_globally_2024, passmore_global_2024, brown_musical_2025, mcbride2025melodypredominates}. A fundamental open question is whether these cultural changes are coupled: when a musical tradition shifts melodically, does it tend to shift rhythmically as well, or can melody and rhythm follow partially independent evolutionary trajectories?

There are reasons to expect dissociation between these two components. Comparative and evolutionary accounts suggest that rhythm-related capacities may have older phylogenetic roots than pitch-related capacities~\citep{patel_comparing_2006, kotz_evolution_2018}. Neuro-cognitive work points to at least partially distinct mechanisms for processing temporal versus pitch structure~\citep{bianco_human_2025, haden_beat_2024}. If melody and rhythm are supported by distinct cognitive and neural systems, cultural transmission could act on them at different rates or along different social boundaries. For instance, one might observe high rhythmic similarity within groups but greater melodic diversification, or vice versa. This leads to a concrete testable prediction. Melody and rhythm should not necessarily co-vary in their degree of cross-cultural diversity, each potentially responding to distinct sociodemographic factors

Existing research, however, often treats music as a monolithic entity or focuses on a single component, making it difficult to test whether melody and rhythm change in tandem or independently~\citep{yurdum_universal_2023, zivic_perceptual_2013, jacoby_extreme_2021}. A recent exception is \cite{mcbride2025information}, who studied co-variation of pitch and duration features using transcribed corpora and found correlated entropy between melodic and rhythmic dimensions arising from the same vocal melody. Our work extends this line of inquiry using audio-derived features and a different operationalization of rhythm---percussive timing rather than vocal rhythm.

Scaling this kind of cross-cultural comparison remains challenging. Traditional approaches rely on manual transcription and ethnographic
annotation~\citep{mehr_universality_2019, lomax_folk_1978, savage_statistical_2015} are often constrained by limited sample size, and it can introduce bias when analysis is filtered through Western-derived musical
concepts~\citep{rice_modeling_2017, clayton_cultural_2003,
daikoku_agreement_2022}. 
Standard analytical frameworks built on equal temperament, tonal centers, or beat isochrony can misrepresent non-Western traditions~\citep{hannon_familiarity_2012, polak_rhythmic_2018, jacoby_universal_2019}. 
Recent machine learning approaches have improved scalability~\citep{castellanos_region-based_2022, mauch_evolution_2015}, but often perpetuate similar biases due to training predominantly on Western music datasets~\citep{lee_globalmood_2025}.

\begin{figure*}[t]
  \begin{center}
    \includegraphics[width=\textwidth]{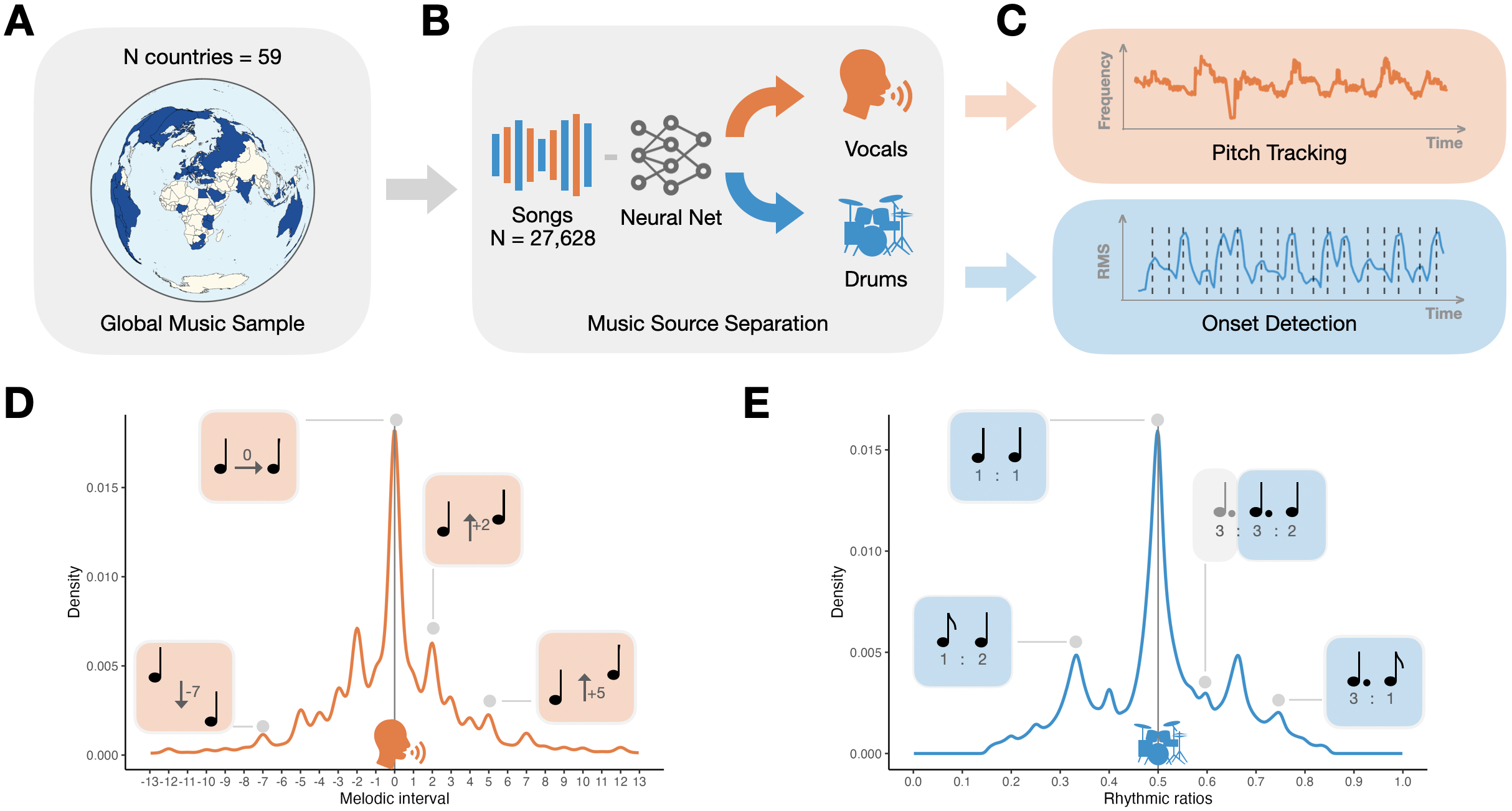}
  \end{center}
  \caption{
    \textbf{Computational pipeline for extracting melodic and rhythmic characteristics from the raw audio.} 
    (\textbf{A}) Global coverage of the dataset comprising 27,628 popular songs sampled from YouTube music charts across 59 countries (`\nameref{subsec:data_collection}' in Methods). (\textbf{B}) Deep learning-based source separation technique decomposes each song into vocal and drums, enabling independent analysis of melody and rhythm (`\nameref{subsubsec:source_separation}' in Methods). (\textbf{C}) Extracted components undergo specialized processing (`\nameref{subsec:feature_extraction}' in Methods): fundamental frequency tracking for vocals (orange) and onset detection for drums (blue). (\textbf{D}) Global distribution of melodic intervals shows known characteristic peaks such as the unison (0), major second ($\pm 2$), and perfect fifth ($\pm 7$), while a characteristic dip at tritone ($\pm 6$). (\textbf{E}) Distribution of rhythmic inter-onset ratios reveals preferences for simple integer relationships: isochrony ($1:1$), duple ($1:2$ / $2:1$), triple ($1:3$ / $3:1$), and $3:2$ / $2:3$ that are elements to forms more complex patterns such as \textit{tresillo} ($3:3:2$). Musical notation illustrates some of these melodic intervals and rhythmic ratios at the distribution peaks.
  }
  \label{fig:pipeline}
\end{figure*}

Here, we introduce a computational pipeline that extracts melodic and rhythmic distributional profiles directly from raw audio at scale (Figure~\ref{fig:pipeline}). At its core, the pipeline uses deep learning source separation~\citep{rouard_hybrid_2023} to decompose each recording into a vocal and a percussive component, then characterizes each through distributions of pitch intervals (melody) and inter-onset timing ratios (rhythm). Crucially, this \textit{within-song} matched measurement means that both components derive from the same recording rather than from different datasets, annotation conventions, or analytical frameworks. By working with distributional features rather than higher-level constructs, the pipeline minimizes the number of analytical commitments such as key, scale, or meter.

We apply this pipeline to a novel dataset of 27,628 country-exclusive popular songs from 59 countries, sampled from YouTube's weekly music charts (Figure~\ref{fig:pipeline}A).
By focusing on songs that chart locally rather than internationally, the dataset captures region-specific musical preferences. We then test whether melody and rhythm exhibit coupled versus independent patterns of cultural variation by examining whether the same sociodemographic boundaries constrain both components.
We here focus on one specific quantity, distributional diversity, as a measure of cross-cultural variation, while noting that other aspects of melodic and rhythmic structure (e.g., entropy, central tendency, sequential dependencies) remain to be investigated.

\section{Methods}
\label{sec:methods}

\subsection{Data collection}\label{subsec:data_collection}
Our dataset consists of weekly Top 100 song charts from YouTube Music (\url{https://charts.youtube.com}) collected between September 2017 and February 2022. The initial dataset contained 1,067,100 entries across the all available 61 countries, comprising 64,945 unique songs. From this full set, we identified songs exclusive to individual countries to capture musical characteristics specific to that country (and not the international mega hits that are popular everywhere). After excluding songs that appeared in multiple countries (26.1\%), we retained 47,995 country-exclusive or `locally niche' songs. To balance computational feasibility with comprehensive coverage, we implemented a 1,500-song limit per country. For countries exceeding this threshold, we randomly sampled 1,500 songs, while for others, we took all existing songs. We excluded countries with fewer than 50 songs to achieve enough statistical power. This removed El Salvador (n=18) and Costa Rica (n=24), resulting in a final set of 59 countries and 27,628 unique songs. 


\subsection{Feature extraction pipeline}\label{subsec:feature_extraction}

\subsubsection{Source separation}\label{subsubsec:source_separation}
We used the Demucs music source separation tool (version 4) to extract vocal singing and drum parts from the raw audio~\citep{rouard_hybrid_2023}. Demucs is a state-of-the-art model capable of separating vocals, drums, bass, and other accompaniment parts from the song. For processing efficiency, we used the `mdx\_q' model variant and segmented each song's raw audio into 1-minute clips taken from random points. Using 32GB GPU and 64 CPU cores in parallel, we processed all 27,628 audio files in approximately 3 days.

\subsubsection{Melodic features}\label{subsubsec:melodic_features}
We tracked fundamental frequency ($f0$) in the source-separated vocal tracks to capture pitch contours and infer the distribution of melodic intervals. $f0$ was extracted from the audio recordings using the `pyin' pitch tracking algorithm implemented in the `librosa' Python package~\citep{mauch_pyin_2014}. This process yields time-stamped frequency values representing the vocal melody.

Next, we computed melodic intervals from the pitch contours by calculating semitone differences between proximal pitch values. Frequencies were first converted to MIDI pitch values using $m(t) = 12 \log_2(f(t)/440) + 69$, where $f(t)$ is the fundamental frequency at time $t$. Then we used a sliding time window, where intervals between pitches were analyzed at varying timescales, ranging from 100ms to 2,000ms, with a step size of 10ms. For a given time lag $\Delta$, the melodic interval at time $t$ was computed as:
$$I(t, \Delta) = m(t) - m(t - \Delta)$$
For instance, with a time window of $\Delta = 100$ms, we computed the difference in semitones between the pitch at each time point and the pitch 100ms earlier; with $\Delta = 2000$ms, we captured slower melodic movements spanning up to two seconds. The minimum and maximum limits of these time windows were chosen based on literature to encompass both rapid ornamental figures and slower melodic gestures typical of vocal music~\citep{london_cognitive_2002}.

To create a continuous representation, we pooled interval values across all time lags (from 100 ms to 2,000 ms in 10 ms steps) into a single set and then applied kernel density estimation (bandwidth = 0.10 semitones), producing one summary distribution per song. This pooling captures melodic relationships at multiple timescales simultaneously. We acknowledge that this approach yields an approximation of traditional pitch-interval distributions, and that formal evaluation against ground-truth transcribed datasets is an important direction for future work.

\begin{figure*}[!htbp]
  \begin{center}
    \includegraphics[width=\textwidth]{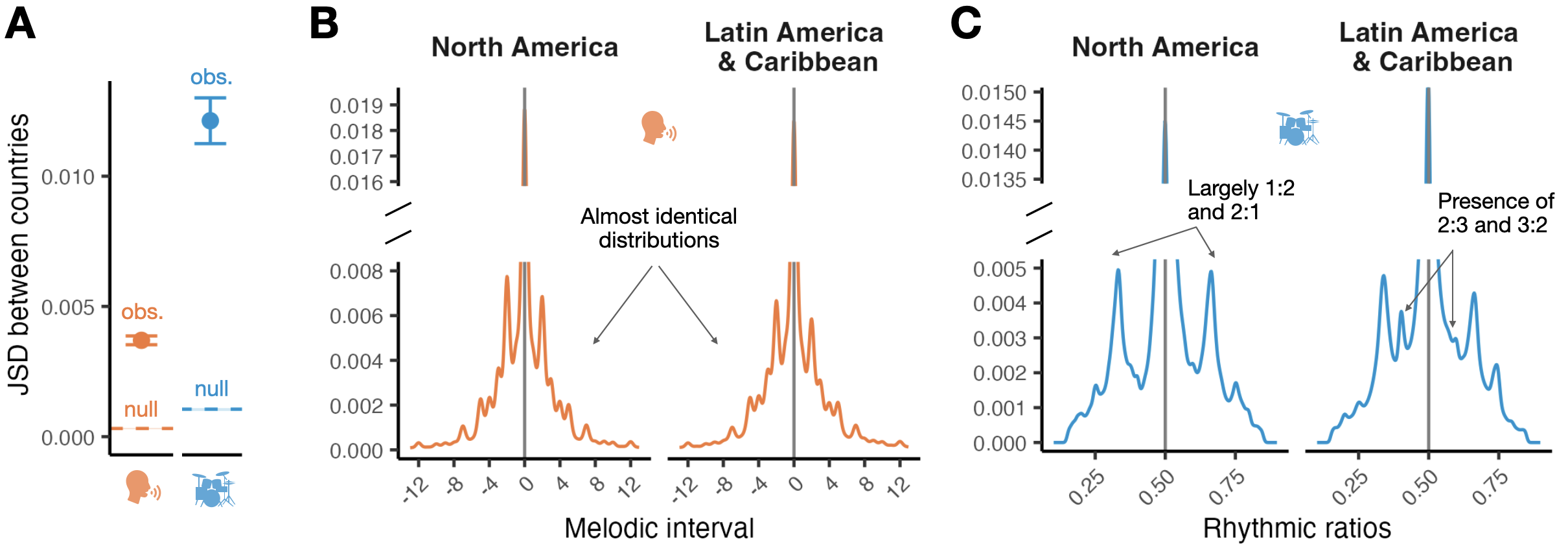}
  \end{center}
  \caption{
    \textbf{Cross-cultural variation in melodic and rhythmic structure.} 
    (\textbf{A}) Observed Jensen-Shannon divergence (JSD) between country-level distributions of melody and rhythm compared against a null model that shuffles country labels across songs while preserving sample sizes. Both domains show significantly greater between-country divergence than expected by chance. Error bars represent 95\% CI across bootstrap iterations. 
    (\textbf{B}) Melodic interval distributions for North America and Latin America \& Caribbean (shown as example regions) exhibit strikingly similar profiles, whereas 
    (\textbf{C}) rhythmic ratio distributions for the same regions show marked differences. Y-axes are truncated to highlight variation beyond the dominant peaks at unison (melodic interval = 0) and isochrony (rhythmic ratio = 0.5).}
  \label{fig:diversity}
\end{figure*}

\subsubsection{Rhythmic features}\label{subsubsec:rhythmic_features}
We tracked the timing patterns of percussive onsets in the source-separated drum tracks and computed inter-onset ratios that capture rhythmic structures independent of tempo. The source separation algorithm isolates the rhythmic component into a single track, combining all percussive elements (e.g., hi-hat and bass drum) into a unified stream with audible and typically separated onsets.

First, we extracted note onset times from the unified audio recordings using an onset detection algorithm implemented in the librosa package. This process yields time-stamped markers indicating the start of each musical event. The method relies on sets of heuristics that track strong peaks on the spectral novelty function, rather than being trained on specific music corpora (which could introduce certain cultural biases).

Second, we computed inter-onset ratios from consecutive onset triplets following the approach of \cite{roeske_categorical_2020}. For each set of three consecutive onsets at times $t_1$, $t_2$, and $t_3$, we calculated the two inter-onset intervals $\text{IOI}_1 = t_2 - t_1$ and $\text{IOI}_2 = t_3 - t_2$, and then computed the normalized ratio:
$$r = \frac{\text{IOI}1}{\text{IOI}_1 + \text{IOI}_2}$$
This ratio represents the proportion of the first interval within a two-interval rhythmic cycle, yielding values between 0 and 1. Ratios below 0.15 or above 0.85 were excluded to filter out extremely uneven subdivisions likely arising from onset detection errors or grace notes. This ratio-based measure is tempo-invariant, enabling direct comparison across pieces with different tempi.

Finally, just as with melodic distribution, we applied kernel density estimation on the inter-onset ratios (bandwidth = 0.005) and normalized the density profile to sum to one. 


\subsection{Socio-demographic factors}\label{subsec:sociodemographic_factors}
We compared melodic and rhythmic diversity within each country with demographic diversity along four dimensions: ethnicity, language, religion, and genetics. Data for ethnic, linguistic, and religious diversity were derived from Fearon's Fractionalization index~\citep{fearon_ethnic_2003}. Fearon's index quantifies a country's diversity as the probability (ranging from 0 to 1 with 1 indicating maximum fragmentation) that two randomly selected individuals belong to different groups. 

Data for genetic diversity were derived from the GeLaTo database~\citep{barbieri_global_2022}, which compiles original data collected by \cite{pemberton_genomic_2012}. We used heterozygosity as our measure of genetic diversity, which measures the proportion of individuals in a given population that carry two different alleles at a given locus.

\section{Results}
\label{sec:results}

\subsection{Global melodic and rhythmic patterns provide face validity}\label{subsec:validation_our_approach}
Before examining cross-cultural variation, we first validate our computational pipeline by aggregating melodic and rhythmic distributions across all songs and countries to produce global summaries (Figures~\ref{fig:pipeline}D\&E). The melodic interval distribution (Figure~\ref{fig:pipeline}D) reproduced known peaks at small intervals and the characteristic tritone dip~\citep{vos1989ascending,huron2006sweet,brown_musical_2025,tenney1988consonance,anglada-tort_large-scale_2023}, while the rhythmic distribution (Figure~\ref{fig:pipeline}E) showed the expected preferences for isochrony and small-integer ratios~\citep{jacoby_commonality_2024, polak_rhythmic_2018, roeske_categorical_2020}. These convergent patterns confirm that the pipeline successfully recovers established regularities in both melodic and rhythmic structure, hence providing face validity for the subsequent cross-cultural analyses.

\subsection{Geolinguistics explains between-country musical similarities}\label{subsec:between_country}
To examine whether melodic and rhythmic patterns demonstrate culturally distinct signatures, we computed Jensen-Shannon divergence (JSD) between all pairs of countries for both melodic intervals and rhythmic ratios, and compared these observed distances to a permutation null model (Figure~\ref{fig:diversity}A). The null model shuffles country labels across songs while preserving sample sizes within each country, allowing direct comparison between observed and chance-level between-country distances. Both melody and rhythm showed significantly greater between-country divergence than expected by chance (melody: $z$ = 10.0, Cohen's $d$ = 1.68, $p$ < 0.001; rhythm: $z$ = 14.0, $d$ = 1.28, $p$ < 0.001), confirming that musical features cluster by country origin rather than arising from identical universal distribution.

We further assessed whether the countries that are geographically closer also share similar melodic and rhythmic characteristics. Using geographical regions defined by the World Bank\footnote{\url{https://www.worldbank.org}}, we found that countries within the same geographic region exhibited significantly more similar musical profiles than countries in different regions. Specifically, we compared the JSD values of all same-region country pairs against all different-region country pairs, yielding moderate effect sizes for both melody (Cohen's d = 0.39, p < 0.001) and rhythm (d = 0.58, p < 0.001). We also tested linguistic proximity by computing Spearman correlations between pairwise musical JSD and pairwise linguistic distance~\citep{gurevich_dataset_2025}. Musically similar countries tend to share more similar spoken languages (melody: $rho$ = 0.26, bootstrapped 95\% CI = [0.04, 0.45], $p$ = 0.016; rhythm: $rho$ = 0.27 [0.10, 0.44], $p$ = 0.002). Together, these expected geolinguistic alignments provide convergent validity that our extracted melodic and rhythmic features capture culturally meaningful variation.

However, regional proximity does not imply uniform similarity across both musical dimensions. As visual examples, Figures~\ref{fig:diversity}B\&C illustrate regional aggregate profiles comparing North America versus Latin America \& the Caribbean, showing that the extracted melodic-interval and rhythmic-ratio distributions yield interpretable, systematically different patterns across regions. For these two regional comparisons, the melodic intervals appear very similar, suggesting no major differences in melodic structure. In contrast, the rhythmic distributions show noticeable differences, namely the prominence of 2:3/3:2 ratios in Latin America \& the Caribbean. These ratios form subcomponents of more complex rhythmic patterns, such as the 3:3:2 \textit{tresillo} \citep{rocamora2018computational,kubik1979angolan}, and have also been reported as significant in human perceptual experiments conducted in this region \citep{jacoby_commonality_2024}.

\subsection{Melodic and rhythmic diversity are independent}\label{subsec:independence_melody_rhythm}
\begin{figure*}[!htbp]
  \begin{center}
    \includegraphics[width=\textwidth]{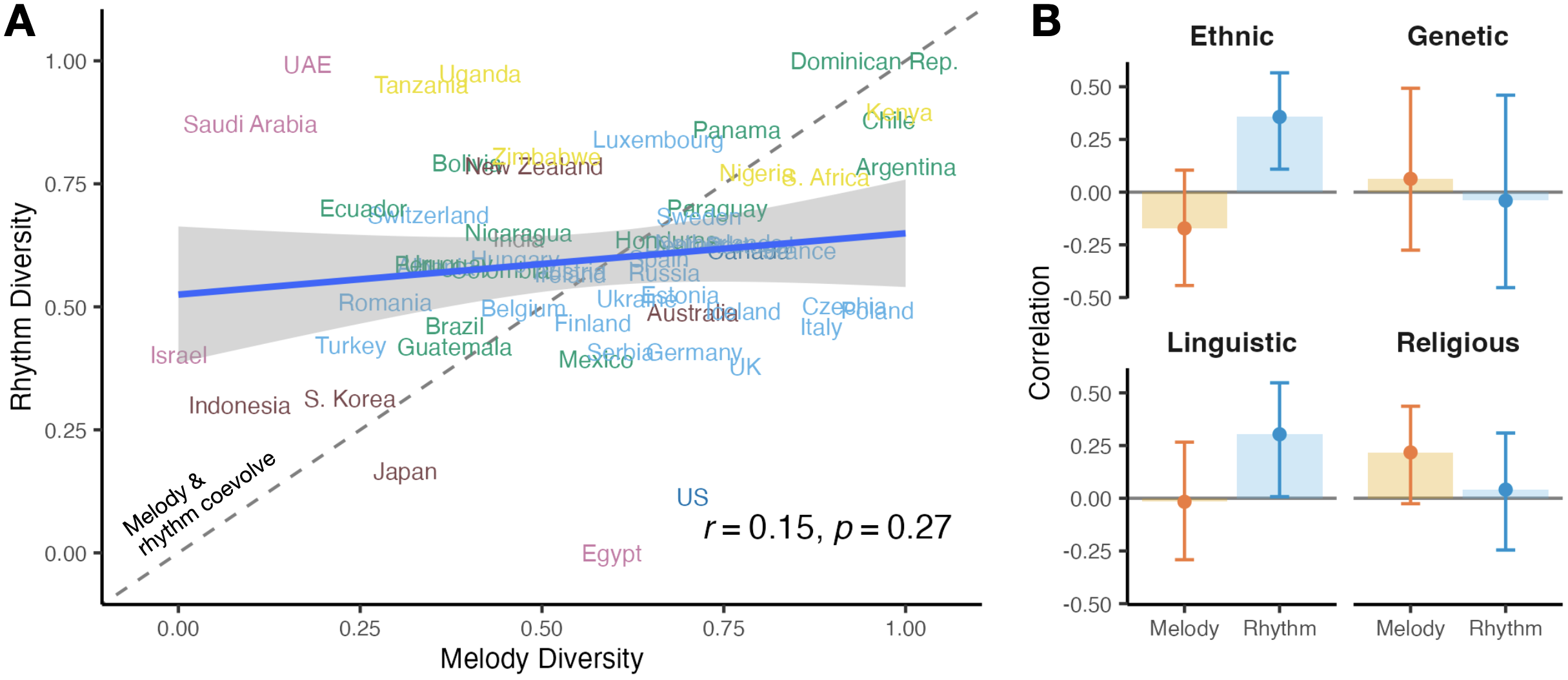}
  \end{center}
  \caption{\textbf{Within-country melodic and rhythmic diversity demonstrate independence.} (\textbf{A}) Scatter plot of melodic diversity versus rhythmic diversity across countries, colored by world region, showing no significant relationship between the two ($p = 0.27$). The dashed diagonal line represents perfect correlation, a theoretical scenario in which the two components could have coevolved, resulting in similar levels of diversities. The observed pattern towards being flat indicates independence between these two components. (\textbf{B}) Association between musical diversity and diversity in population demographics (`\nameref{subsec:sociodemographic_factors}' in Methods), showing countries with diverse mix of ethnic groups and language tend to have more diverse rhythm but not melody. Error bars indicate bootstrapped 95\% CI of the mean.}
  \label{fig:socio-demographic}
\end{figure*}

Having established that the pipeline captures meaningful variance of musical characteristics, we now directly tested whether heterogeneity in one component predicts heterogeneity in the other (coupled evolution) or whether they vary independently. We quantified the diversity of melodic and rhythmic profiles within each country by computing the JSD between all pairs of songs, summarizing by the median (robust to skew and outliers), and normalizing values to range between 0 and 1, where 1 indicates maximum diversity.

Figure~\ref{fig:socio-demographic}A plots melodic diversity against rhythmic diversity across all 59 countries and shows no significant association ($p = 0.27$). Under a coupled-change account, countries with high melodic diversity should also exhibit high rhythmic diversity, yielding a positive relationship aligned with the diagonal. Instead, the relationship is effectively flat, indicating that melodic and rhythmic diversity vary independently across cultures. To ensure this pattern is not driven by regional clustering (i.e., Simpson's paradox), we performed partial correlation by removing regional means, which revealed the same outcome ($p = 0.41$).

Regional structure nevertheless revealed substantial differences in where countries fall in this two-dimensional space. Sub-Saharan African countries cluster toward high rhythmic diversity (mean regional diversity $D$ = 0.86, 95\% CI = [0.76, 0.96]). Europe and Central Asia occupy an intermediate range for both melodic and rhythmic diversity (melody = 0.64 [0.56, 0.71]; rhythm = 0.55 [0.51, 0.59]). Middle East and North Africa show notably low melodic diversity (0.22 [$-0.20$, 0.64]), while East Asia and Pacific (particularly Japan, Korea, and Indonesia) appear relatively homogeneous in both dimensions (melody = 0.37 [0.067, 0.66]; rhythm = 0.41 [0.12, 0.71]).

Together, these results indicate that melodic and rhythmic diversity are unevenly distributed across regions, and that high diversity in one component does not imply high diversity in the other. In particular, Sub-Saharan Africa stands out as especially diverse in rhythmic patterns, consistent with ethnomusicological accounts emphasizing the diversity of rhythm in many African musical traditions~\citep{polak_mali_2025}.

\subsection{Independent socio-demographic constraints}\label{subsec:sociodemographic_constraints}

If melody and rhythm evolve independently, they may also respond to different socio-demographic constraints. We tested this by examining whether within-country musical diversity reflects broader patterns of demographic diversity across ethnic, genetic, linguistic, and religious dimensions (`\nameref{subsec:sociodemographic_factors}' in Methods). Rhythmic diversity showed positive associations with both ethnic diversity ($r$ = 0.36, $p$ = 0.007, Benjamini-Hochberg $p_{\text{adj}}$ = 0.054) and linguistic diversity ($r$ = 0.30, $p$ = 0.044, $p_{\text{adj}}$ = 0.18), suggesting that countries with more heterogeneous populations tend to exhibit greater variety in their rhythmic patterns. In contrast, melodic diversity showed no significant association with any demographic diversity measure (all $p$ > 0.08, all $p_{\text{adj}}$ > 0.21). Consistent with recent findings by \cite{passmore_global_2024} showing weak links between musical and genetic similarities, neither melody nor rhythm demonstrated associations with genetic diversity. 
Together, these correlations possibly suggest that rhythmic diversity weakly aligns with ethnolinguistic heterogeneity, consistent with findings of perceptual enculturation and niche construction in rhythm~\citep{polak2026cultureneuro}, whereas melodic diversity remains decoupled from these demographic dimensions.

\section{Discussion}
\label{sec:discussion}
Our computational pipeline enabled the large-scale, within-song comparison of melodic and rhythmic variation across cultures. By extracting both components from the same recordings using source separation, we bypassed the scalability limitations and potential biases inherent in manual annotation systems. This methodological advance revealed cross-cultural patterns that would otherwise be difficult to detect through traditional approaches. 

We found that melody and rhythm were not significantly correlated in their within-country diversity. For instance, Sub-Saharan Africa's exceptionally high rhythmic diversity combined with moderate melodic diversity, or the near-identical melodic profiles but divergent rhythmic distributions between North America and Latin America \& the Caribbean, exemplify how cultural exchange can homogenize one dimension while preserving distinctiveness in another. These patterns suggest that melody and rhythm are not merely different features of a monolithic `musical style', but rather partially independent systems responding to different evolutionary and cultural pressures.

These results complement the findings of \cite{mcbride2025information}, who operationalized both melody and rhythm from the vocal line itself and found correlated entropy between pitch and duration dimensions, suggesting co-evolution. Our finding that vocal melody and percussive rhythm vary independently is not contradictory. The rhythm extracted from drums and the rhythm inherent in vocal phrasing likely reflect different levels of musical organization. In the case of \cite{mcbride2025information}, vocal pitch and vocal rhythm are produced by the same "instrument" (the singer) and constrained by the same physiology such as breathing, whereby we would expect some coupling between the two. Percussive rhythm, by contrast, involves separate performers, instruments, and transmission pathways, allowing it to diverge from melodic structure.

Our results point to potential differential mechanisms of cultural transmission. Rhythm appears more tightly coupled to local population structure, with rhythmic diversity tracking ethnolinguistic heterogeneity, a pattern consistent with rhythm serving as a core constituent of cultural practice~\citep{polak2026cultureneuro} and marker of group identity and social cohesion~\citep{savage_music_2020, brown_correlations_2014}. In contrast, melodic diversity shows no such demographic associations, suggesting that melody may more easily diffuse across cultural boundaries and be more adaptable to change than rhythm~\citep{bomin_evolution_2016}.

\subsection{Limitations and Future Works}
\label{subsec:limitations_future_works}
The reliance on YouTube chart data may underrepresent traditional and non-commercial music, particularly in undersampled regions like Sub-Saharan Africa. Future work should incorporate traditional music repositories to ensure broader cultural coverage~\citep{jacoby_cross-cultural_2020}. Moreover, like many applications in automatic music analysis, source separation algorithms are primarily trained on Western popular music and may carry cultural biases in distinguishing vocal and drum tracks. For instance, the same algorithm would likely fail to extract timpani lines from Western classical orchestral music. A systematic evaluation of source-separation quality across genres and decades would strengthen confidence in the pipeline; for instance, running a human listener evaluation experiment across a diverse selection of music~\citep{lee_cross-cultural_2021, lee_globalmood_2025}.

Our operational definitions are necessarily selective: we define melody as pitch trajectory (voice only) and rhythm as inter-onset intervals (drums only), thereby excluding contributions from other instruments and overlooking how voice conveys rhythm, drums convey melody, and ensemble interplay creates emergent melodic-rhythmic structures. Similarly, we do not capture how melody is shaped by timbre and instrumentation, or how rhythm is constructed from event duration and pitch-timbre contrasts. While these choices reflect practical constraints, future work could systematically compare alternative operationalization. We also note that melodic pitch-interval distributions span a higher-dimensional space than the bounded rhythmic ratio distributions (which fall between 0 and 1), which may make melodic diversity inherently noisier to estimate from limited samples. Future work should assess the sensitivity of diversity estimates to sample size and alternative summary statistics (e.g., entropy).

Finally, our cross-sectional design captures contemporary patterns but cannot establish causal relationships between socio-demographic factors and musical features. The observed differential associations that, rhythm is linked to ethnolinguistic heterogeneity while melody does not, may reflect independent evolutionary trajectories, but future work using longitudinal datasets or natural experiments~\citep{lee_mechanisms_2025, anglada-tort_here_2023} are needed to test causal hypotheses and trace historical change.

\section{Code and Data Availability}
\label{sec:code_data_availability}
Audio obtained for research analysis purposes cannot be redistributed and thus not made available. Instead, we make available the full metadata of the sampled songs in the study, along with core analysis scripts, and aggregated distributions of melody and rhythm. Publicly available at \url{https://github.com/harin-git/mel-rhy}.

\section{Acknowledgments}
\label{sec:acknowledgments}
We greatly appreciate the feedback from the reviewers, including John McBride, who provided detailed and constructive comments that substantially improved this manuscript.

\printbibliography

\end{document}